# Security implications of user non-compliance behavior to software updates: A risk assessment study

Mahzabin Tamanna [*], Mohd Anwar, Joseph D.W. Stephens

*North Carolina A&T State University, United States of America*

## A R T I C L E  I N F O

## A B S T R A C T

Software updates are essential to enhance security, fix bugs, and add better features to the existing software. While some users accept software updates, non-compliance remains a widespread issue. End users' systems remain vulnerable to security threats when security updates are not installed or are installed with a delay. Despite research efforts, users' noncompliance behavior with software updates is still prevalent. In this study, we explored how psychological factors influence users' perception and behavior toward software updates. In addition, we investigated how information about potential vulnerabilities and risk scores influence their behavior. Next, we proposed a model that utilizes attributes from the National Vulnerability Database (NVD) to effectively assess the overall risk score associated with delaying software updates. Next, we conducted a user study with Windows OS users, showing that providing a risk score for not updating their systems and information about vulnerabilities significantly increased users' willingness to update their systems. Additionally, we examined the influence of demographic factor, gender, on users' decision-making regarding software updates. Our results show no statistically significant difference in male and female users' responses in terms of concerns about securing their system. The implications of this study are relevant for software developers and manufacturers as they can use this information to design more effective software update notification messages. The communication of the potential risks and their corresponding risk scores may motivate users to take action and update their systems in a timely manner, which can ultimately improve the overall security of the system.

## 1. Introduction

A software update involves making adjustments to improve or fix issues with the software. The updates can range from minor changes to significant enhancements or the addition of new features. These updates are important not only for fixing vulnerabilities and bugs but also for maintaining the security of the software [1]. One of the most important features of any modern security system is its capacity for releasing effective and safe software upgrades [2]. Software update mechanisms try to ensure accessibility, efficiency, robustness, and expandable distribution of software updates to facilitate the timely application of security patches [3–6]. In 2021, a vulnerability known as Log4j or Log4Shell impacted approximately three billion computer systems and applications [7]. This attack was considered "critical", with a severity score of 10 according to the National Vulnerability Database (NVD) [8]. According to the experts, this attack could have been prevented if the available software update had been applied [9]. As per a study conducted in 2022, many security breaches occur because of uninstalled updates for vulnerabilities that were available [10]. In general, vulnerable software is usually targeted by cyber attackers, especially if that software has widespread use and a large number of users, such as Microsoft Office, Adobe Acrobat, and Internet Explorer [11]. While technical remedies for security concerns, such as releasing software patches, hold significance, improving human awareness toward security practices is indispensable for achieving cyber safety. The security of a computer system often relies on how users understand, behave, and make security-related decisions [12–14]. Failing to install updates for identified vulnerabilities can lead to severe security breaches. Previous work suggested that many users do not perceive all updates as equally significant or prioritize them appropriately [15]. This can leave systems exposed to potential threats and undermine the effectiveness of security measures.

Prior survey-based studies delved into users' behavioral studies and found users' unfavorable behaviors impact both individual security [16] and organizations' security stance [17]. In organizations, approximately 27% of data breaches are caused by not adopting common security and privacy measures by the end users [18]. Similarly, end users often fail to adopt common security and privacy measures [19,20]. One of the main reasons for security violations is a failure to apply the patch for a known vulnerability, as most of the exploitation occurs in systems that are not updated [21]. Microsoft reported that most of its customers are breached via vulnerabilities that had patches released years ago. This indicates users' non-compliance behavior toward applying patches [21]. According a survey prefomed by to Voke media, about 80% of companies with a data breach or failed audit could have prevented that by patching on time or doing configuration updates [22]. Similarly, 84% of the companies have high-risk vulnerabilities on their external networks; more than half of those vulnerabilities could have been simply removed just by installing the update [23]. Once vulnerabilities are discovered, developers generally release an update or patch as quickly as possible. On average, patch release time for a vulnerability ranges between 23 to 40 days [24,25]. Arora et al. [26] showed, after the discovery of the vulnerability, the white-hat developers generally take 30 to 45 days to make the patch available. In this timeline, if the vulnerability is discovered by the black-hat community (i.e., hackers with malicious intent), then the vulnerability could be exploited within zero days. Frei [27] found that 78% of exploitations take place within a day, and 94% do so within 30 days of the public disclosure day. However, studies suggest that the gap between public disclosure and exploitation is decreasing, and the exploitation process now takes around five days [28]. Furthermore, prior research has also shown that there are differences in user behaviors based on gender [29]. Researchers found that female users exhibited lower levels of security and privacy behaviors than male users in 40% of cases. Also, female users showed less interest in adopting technical skills than male users [30]. Hence, it is clear that software security not only depends on the timely release of software updates but also on users' compliance in updating their systems to prevent attacks [31,32].

As the previous works present, people are negligent when it comes to installing updates and prefer to delay the process, while early updating applications could make the system secure and protect it from unwanted attacks. Our study is focused on analyzing how delaying software updates could increase cyber risk and investigating software update behaviors of users when they have information about vulnerabilities and risk scores. Additionally, in the field of psychology, research has explored attitudes, including their correlation, antecedents and consequences, and correlation with intentions and behavior [33,34]. To gain a better understanding of users' security attitudes, we analyzed their attitudes toward cybersecurity.

In this study, we investigated end-users non-compliance behavior and their perspectives toward software updates. We proposed a framework to assess how delaying updating software can increase security risk. Next, we developed a questionnaire combining skills, awareness, experience, and knowledge-based questions. We designed our questionnaire to evaluate users' behavior and changes in their decision-making process before and after having proper vulnerability and security risk-related information. Further, using a combination of factors and statistical analysis, we identified which factors increase users' awareness and influence them to update software to secure their system. Additionally, we extended our research to determine if male and female end users' behavior and perception differ when they have the same security-related information. With this extension, we looked into the discrepancies in gender-based security behavior. To achieve the above-stated research objectives, we focused on the following research questions:

- RQ1: How do users' cognitive states affect their adoption of software updates?
- RQ2: To what extent does the vulnerability and risk score information improve users' software update compliance behavior?
- RQ3: What difference does gender make in software update decision-making?

This paper has been organized as follows. Section 2 describes the background of CVSS. Section 3 discusses the related work on users' software update behaviors and risk estimation from software update non-compliance. Section 4 presents the methodologies for risk-score assessment associated with software update delays and survey study. Section 5 explains the analysis and results. Finally, Section 6 concludes the paper, and Section 7 discusses the limitations of our study and future works.

## 2. Background

**Common Vulnerability Scoring System (CVSS)**

CVSS is an open framework that aims to offer a global and software-independent rating of all known and recorded vulnerabilities. The CVSS has two versions, version 2.0 and version 3.0. The framework is developed and maintained by the Forum of Incident Response and Security Teams (FIRST) [35], a US-based non-profit organization that serves to help security problems worldwide. More than 33,000 vulnerabilities have been recorded in the National Vulnerability Database (NVD) in 2024. FIRST provides the CVSS score for each vulnerability recorded in the NVD [36]

CVSS has three metric groups, base, temporal, and environmental, to quantify the severity risk level [37]. The score is measured on a decimal number scale [0.0, 10.0], where scales are labeled as [Low, Medium, High]. Though each group generates its numeric score, CVSS only considers the base score to evaluate the risk. In contrast, the other two metrics, temporal and environmental, vary based on the factors related to an organization that uses the computer system and how vulnerability may change over time. The base group evaluates the intrinsic attributes of any individual vulnerability by using two sub-metric groups: (i) exploitability value and (ii) impact value. The exploitability is composed of an access vector (AV) ("reflects how the vulnerability is exploited in terms of local, adjacent network or networks"), access complexity (AC) ("measures the complexity of the attack required to exploit the vulnerability once an attacker has gained access to the target system"), and authentication (Au) ("measures the number of times an attacker must authenticate to a target in order to exploit a vulnerability"). The impact value evaluates the vulnerability's potential impact on confidentiality (C), integrity (I), and availability (A). The impact on confidentiality, integrity, and availability can affect the system as none, partial or complete. The temporal metrics group utilizes the dynamic aspects of a vulnerability that might change over time. The temporal metric score uses three attributes, which are exploitability tools & techniques (E), remediation level (RL), and report confidence (RC). The use of exploit codes or techniques can make a system vulnerable to attacks. An essential consideration for prioritizing is the vulnerability's repair or remediation level. The corresponding patch may not be available when a vulnerability is first disclosed. As the official and proper remediation becomes available, the severity of the vulnerability goes downwards. Report of confidence (RC) gauges the level of assurance on the presence of vulnerability and the reliability of the available technical information. The environmental metrics capture the characteristics of a vulnerability that are associated with a user's environment and evaluate three elements: (i) collateral damage potential (CDP), (ii) target distribution (TD), and (iii) security requirements (CR, IR, AR). The collateral damage potential measures the potential for loss of life or physical assets through damage or theft of property or equipment. Target distribution is an environment-specific indicator to estimate the percentage of systems that could be affected by the vulnerability. The security requirements are confidentiality (CR), integrity (IR), and availability (AR). By using metrics, the analyst can adjust the CVSS score according to the importance of the affected IT asset to the user's organization. More details about the CVSS metrics and scores are given in Table 1. The following Table 2 represents the risk severity levels and the score range.

**Table 1**
CVSS metric groups and values.
*Source:* Adopted from Mell et al. [37]

| CVSS metric group | Metrics | Metrics levels | Level value |
|---|---|---|---|
| Base metrics | Access Vector (AV) | Local (L) | 0.395 |
| | | Adjacent Network (A) | 0.646 |
| | | Network (N) | 1.000 |
| | Access Complexity (AC) | High (H) | 0.350 |
| | | Medium (M) | 0.610 |
| | | Low (L) | 0.710 |
| | Authentication (Au) | Multiple (M) | 0.450 |
| | | Single (S) | 0.560 |
| | | None (N) | 0.704 |
| | Confidentiality Impact (ConfImpact) | None (N) | 0.000 |
| | | Partial (P) | 0.275 |
| | | Complete (C) | 0.660 |
| | Integrity Impact (IntegImpact) | None (N) | 0.000 |
| | | Partial (P) | 0.275 |
| | | Complete (C) | 0.660 |
| | Availability Impact (AvailImpact) | None (N) | 0.000 |
| | | Partial (P) | 0.275 |
| | | Complete (C) | 0.660 |
| Temporal metrics | Exploitability (E) | Unproven (U) | 0.850 |
| | | Proof-of-Concept (POC) | 0.900 |
| | | Functional (F) | 0.950 |
| | | High (H) | 1.000 |
| | | Not Defined | 1.000 |
| | Remediation Level (RL) | Official Fix (OF) | 0.870 |
| | | Temporary Fix (TF) | 0.900 |
| | | Workaround (W) | 0.950 |
| | | Unavailable(U) | 1.000 |
| | | Not Defined (ND) | 1.000 |
| | Report Confidence (RC) | Unconfirmed (UC) | 0.900 |
| | | Uncorroborated (UR) | 0.950 |
| | | Confirmed (C) | 1.000 |
| | | Not Defined (ND) | 1.000 |
| | | Unavailable (U) | 1.000 |
| | | Not Defined (ND) | 1.000 |
| Environmental metrics | Collateral Damage Potential (CDP) | None (N) | 0.000 |
| | | Low (L) | 0.100 |
| | | Low-Medium (LM) | 0.300 |
| | | Medium-High (MH) | 0.400 |
| | | High (H) | 0.500 |
| | Target Distribution (TD) | None (N) | 0.000 |
| | | Low (L) | 0.250 |
| | | Medium (M) | 0.750 |
| | | High (H) | 1.000 |
| | | Not Defined (ND) | 1.000 |
| | Security Requirements (CR, IR, AR) | Low (L) | 0.500 |
| | | Medium (M) | 1.000 |
| | | High (H) | 1.510 |
| | | Not Defined (ND) | 1.000 |

**Table 2**
Risk-Score severity rating scale.
*Source:* Adopted from Mell et al. [21].

| Severity levels | CVSS score range |
|---|---|
| None | 0 |
| Low | 0.1–3.9 |
| Medium | 4.0–6.9 |
| High | 7.0–10 |
| Critical | 9.0–10 |

## 3. Related work

### 3.1. Users' software update behavior

Users' compliance behavior and attitude play an important role in increasing security [12,38–41]. Previous works suggested that increased user involvement can result in better security [42]. However, studies have also shown that users often neglect and delay software updates [43,44]. Software update reluctance is largely caused by the update representation or the user's prior inconvenient experience with software updates, which raises the suspicion that an update may introduce new problems [45,46]. Vaniea et al. [45] found that users' past negative experiences, such as surprise UI changes, cumulative updates, and forced reboots, caused an inclination toward noncompliance with software update messages. Mathur and Chetty [47] analyzed users' attitudes toward software updates and found users who do not update their applications are more likely to have unpleasant experiences and are prone to taking risks. Similarly, a lack of proper awareness and information [48,49] also leads to non-compliance. Numerous frameworks [50–52] and methods [53,54] have been proposed to enhance users' cybersecurity awareness. Vaniea and Rashidi [55] The author of a study has identified six stages that users go through when deciding whether or not to comply with pop-up software update messages. These stages are awareness, decision-making, preparation, installation, troubleshooting, and post-installation. Each of these stages affects users' willingness to comply with the software update message. The author recommends several changes to improve results, such as providing more information about the update, including the estimated installation time, and making resources available during installation. Another group of researchers, Fagan et al. [56], studied how users' experiences and beliefs about software updates influenced their proclivity to ignore the update alerts. After analyzing the data, they found several issues connecting users' experiences and beliefs with noncompliance to updates, noting that 28% of the time, users simply ignore software update messages. In addition, annoyance, confusion, and lack of understanding contributed to some users' noncompliance behavior. Users also mentioned that sometimes they deemed the software update messages fake or unimportant and might negatively affect the usability of their devices, so they simply avoided the software update. Additionally, prior researchers [16,57–60] constructed models to examine how users cognitively perceive threats to computer security. However, understanding how users perceive and respond to security threats is still incomplete. Based on research conducted by Dourish et al. [61], it has been found that users often assign the responsibility of security to individuals whom they believe to be more knowledgeable, such as friends or family members. Furthermore, Ion et al. [62] have found that the security advice given by non-experts and experts is unlikely to match. Another study showed that not only non-expert end users but even system administrators face difficulties and are reluctant to process software updates [63]. Additionally, in terms of platform, this negligence behavior is not only limited to computer platform [64] but also extends to Android [65], smartphones [66], and smart homes [67]. One possible reason for this hesitation could be the disruptive nature of software updates, which can take a long time to complete and interrupt ongoing tasks [66]. The release of a software update is only the first step. Ideally, all users should install the update before attackers have a chance to exploit the vulnerability. However, users often delay the update process significantly, and the attackers strike quickly to take advantage of this weakness. One possible solution could be keeping users out of the loop of software updates by implementing automated updates [68,69] or updating silently [70]. However, users often prefer managing updates manually for various reasons [71]. This preference is typically driven by a desire to control the timing and selection of updates, gain a better understanding of update details before installation, and selectively apply updates to specific apps. Additionally, some users choose manual updates to avoid potential impacts on system performance [72].

Mathur and Chetty [47] found that users who opt to disable automatic updates may have a lower risk tolerance and less trust in applications. Moreover, users prefer to have information about the update to assess whether they want the changes or not. Thus, In 2018, Farik et al. [73] recommended that developers should clearly communicate the importance of updates and improve notification messages. Additionally, user-centered solutions such as providing more information and designing better notifications have been suggested to improve compliance rates. A study [47] proposed providing more information, while Tian et al. [74] emphasized designing better notifications.

*3.2. Differences in cybersecurity behavior based on gender*

Numerous studies have delved into the divergence in security behaviors among organizational employees, differentiating between males and females in the context of information security and privacy practices [75–78]. Additionally, other studies have highlighted discernible gender differences in the utilization of technology [79,80]. Two studies have shown that women are more worried about the use of technology [81,82]. Another study found that women tend to show greater enthusiasm for maintaining software than men but often encounter more difficulties [83]. Some female users also showed a lack of experience in information technology and knowledge of using computers [82]. Thompson and Brindley (2020) discovered that men are more vigilant about privacy and disclosing their information on social media than women. However, women showed higher levels of security and privacy concerns [84–86] and better awareness in terms of security [87]. Several academic studies have reported that women have a greater susceptibility to phishing attacks [75,88] but poorer password behavior [77], and a lower likelihood of adopting privacy-protecting behaviors [89]. These findings also indicate that gender can be a factor in cybersecurity behavior differences. Therefore, the possible gender differences in software update behavior need to be explored.

*3.3. Security risk estimation*

Defining the metrics or standards of measures is the first step in evaluating a system's security risk. There are several approaches available to estimate the risk level for a discrete vulnerability [51,90]. However, very few approaches have been discussed for evaluating the severity score for vulnerability because of user non-compliance behavior with software updates, especially for delaying the update.

Tripathi and Singh [90] proposed security metrics to prioritize vulnerability categories based on the CVSS score. The authors re-evaluated the security score by unifying it with updated availability and vulnerability age to help with more accurate vulnerability categorization. Houmb & Franqueira [91] proposed methodologies to calculate the frequency and impact of potential misuse in a wired network by utilizing CVSS metrics. Based on the resulting impact, identified vulnerabilities are assigned service levels. Additionally, using Markov analysis, the authors analyzed the transitions between different operational service levels to estimate the overall risk associated with the Target of Evaluation (ToE). Joh and Malay [28] used a stochastic risk evaluation model and CVSS metrics to discuss the life cycle of a vulnerability and proposed an approach to calculate the likelihood of an adverse event. CVSS metrics were employed to measure the vulnerability risks and have been leveraged in different areas of study as well [92]. Aksu et al. [93] used CVSS metrics and proposed low-level security metrics (e.g., User Detection) to specifically assess the risk of IT systems. Moreover, using CVSS data, Lenkala et al. [94] proposed a framework for risk associated with cloud carriers and compared the security risk between cloud users and cloud providers. Results reveal that the security metrics of the cloud carrier can differ significantly based on the cloud provider they are associated with, which indicates the need for comparison of frameworks for cloud carriers. Lastly, Sawilla et al. [95] used attack complexity and exploit availability from the CVSS to prioritize a vulnerability from the perspective of attackers for the purpose of protecting critical network assets.

While previous studies have examined human behavior and security risk assessment in software updates, there remains a gap in understanding how users respond when informed about the severity of security risks. Some research has investigated the effectiveness of different representations of updated information, but to the best of our knowledge, no study has specifically explored how providing users with detailed risk information influences their behavior. Our work aims to close this gap by offering new insights into user behavior regarding software updates when security risk information is available by enhancing our understanding of user motivations and decision-making in the update process.

**4. Methodology**

To answer our research questions, we followed comprehensive steps of the methodology. Our methodology consists of five phases. In this section, we presented and discussed each step in detail below.

*4.1. Selection of software*

The first step of the research methodology was to identify several software in order to select the most used software by users. To accomplish this, we limited our scope to the Windows Operating System (OS) only because Microsoft is one of the most popular OS platforms among users and has a wide selection of available software. In order to gather a comprehensive list of commonly utilized software applications among Windows operating system users, a collection of the best available free software offerings for the year 2021 was procured from PC Magazine.[1]

**Table 3**
List of selected software for software update study.

| Category | Software |
| --- | --- |
| Antivirus | Kaspersky |
| Web Browser | Google Chrome |
|  | Mozilla Firefox |
| Conferencing | Zoom |
|  | Skype |
| Multimedia | VLC Media Player |
| Pdf Reader | Adobe Acrobat Reader DC |
| Photoshop | Adobe Photoshop |
| Word Processing | Microsoft Office |
| Operating System | Windows OS |
| Remote Access | TeamViewer |
| Social Media | YouTube |
|  | Facebook |

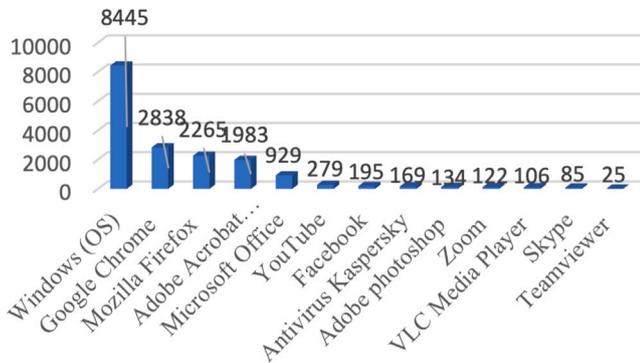

**Fig. 1.** Number of vulnerabilities for each selected software.

This included more than 60 software labeled as "best" based on availability and number of users criteria. Next, we designed and conducted a survey to identify the most popular software from the initial list of 60 software from various categories (e.g., web browser, multimedia, PDF reader). We conducted the survey within the authors' institution and collected data from 63 participants. Participants were asked to rank the software based on their use. Software from different categories, such as browsers, conferencing, and multimedia, was selected. Finally, this study identified a list of 13 software for further analysis. Table 3 lists the selected software along with the category of the software.

### 4.2. Software vulnerabilities and risk score

After selecting the software for the study, we collected the total number of vulnerabilities and risk score (severity score) recorded in the National Vulnerability Database (NVD) [8] (as of 2022). Using the NVD's vulnerability database, we searched for each software listed in Table 3 to collect relevant vulnerability information. This process involved using software names as keywords to retrieve data, which included the total number of vulnerabilities for each software. For each vulnerability, we collected its unique Common Vulnerabilities and Exposures (CVE) ID, a summary of the vulnerability, its publication date, and its CVSS (severity score). Fig. 1 represents the number of total recorded vulnerabilities per selected software.

---

[1] https://www.pcmag.com/picks/best-free-software.

### 4.3. Metrics used to develop the risk-score equation

A vulnerability is more likely to be exploited by attackers if the risk score of the vulnerability is higher. CVSS provides a detailed elaboration of any recorded vulnerability; thus, we leveraged the CVSS to estimate the proposed risk score. As shown in Fig. 2, our risk scoring model utilizes a subset of the Common Vulnerability Scoring System (CVSS) attributes provided by the National Vulnerability Database (NVD), specifically from the Base and Temporal metric groups. From the Base group, we used the following attributes: Access Vector (AV), Access Complexity (AC), Authentication (Au), Confidentiality Impact (C), Integrity Impact (I), and Availability Impact (A). From the Temporal group, we included: Exploitability (E) and Remediation Level (RL) These attributes were selected because they are widely accepted in prior work as key determinants of the exploitability and severity of a vulnerability [37,96–98]. The base metrics capture the intrinsic properties of the vulnerability, while temporal metrics adjust the severity based on exploitation maturity and remediation availability, making them relevant to a model that simulates delayed patch behavior.

### 4.4. Equation formation

The change in risk severity level considering the effects of time. In order to estimate the level of risk associated with any given vulnerabilities, it was necessary to consider an assessment of several important factors. Specifically, we considered the base score, temporal score, and estimated time required for users to apply any available patch. The base score represents the inherent level of risk associated with a given vulnerability, which is subsequently adjusted by applying a temporal factor. The final risk score is then calculated, taking into account the impact of time on the severity of the risk. This comprehensive process enables us to provide a reliable and accurate assessment of the risk level associated with any potential vulnerabilities and helps us take appropriate measures to mitigate any potential risks. Fig. 2 represents the metrics and attributes used in the formulation of the equation to assess the risk score.

#### 4.4.1. Calculation of the base score

CVSS utilizes the base metric group to determine the risk level. The base metric group has two sub-metric groups: (Base_Exploitability) and (Base_Impact). Each of these two sub-metric groups has nine different metric values. As a result, 81 different categories of risk can be calculated based on the vulnerability characteristics. The complete process of estimating the base score includes three steps with six attributes and delivers a severity score in the range from 0 (no severity) to 10 (extreme severity). Following the CVSS equation derived by NVD [8]. First, we calculated the (Base_Exploitability) of the vulnerability using three base metrics: access vector (AV), attack complexity (AC), and authentication instances (AU). Eq. (1) represents the formula to determine the exploitation value *(Base_Exploitability)*. Next, the (Base_Impact) score of the vulnerability is determined using the confidentiality impact (ConfImpact), integrity impact (IntegImpact), and availability impact (AvailImpact) metrics. The method for calculating base impact *(Base_Impact)* is depicted in Eq. (2). [8].

$$Base\_Exploitability = 20 * AV * AC * AU \tag{1}$$

$$\begin{aligned}Base\_Impact = 10.41 \times (1 - (1 - \text{ConfImpact}) \\ \times (1 - \text{IntegImpact}) \times (1 - \text{AvailImpact}))\end{aligned} \tag{2}$$

Let $BS_i$ be the base score for the $i$th vulnerability. Next, base scores for CVSS are calculated by considering the weighting factor and the scores

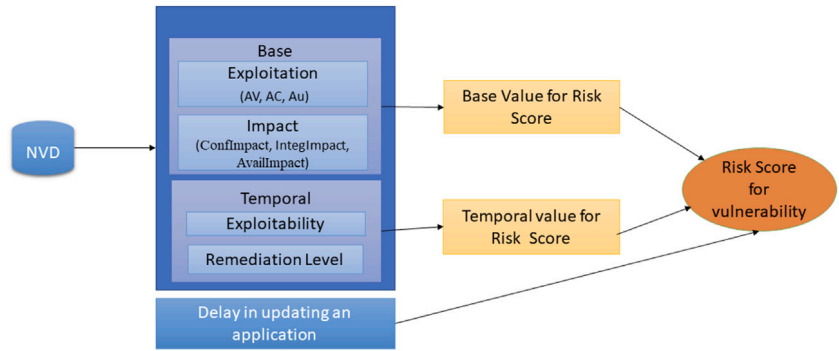

**Fig. 2.** Attributes for estimating risk score.

of the two sub-metrics (exploitability and impact) groups, as shown in Eq. (3).

$$\text{BaseScore (BS}_i) = \text{round\_to\_1\_decimal} \left(0.6 \times B_{\text{Impact}} \right.$$
$$\left. + 0.4 \times B_{\text{Exploitability}} - 1.5 \times f(\text{Impact})\right) \quad (3)$$

where,

$$f(\text{Impact}) = \begin{cases} 0 & \text{if Impact} = 0 \\ 1.179 & \text{otherwise} \end{cases}$$

From the described equations, we calculate the base risk score for each vulnerability.

*4.4.2. Calculation of temporal score*

According to Mell et al. [21], as the threat caused by vulnerability evolves over time, the risk level should be adjusted according to this change. The temporal factor is calculated to reflect this change. Two aspects have been considered while estimating the temporal score: (1) the availability of patches and (2) the state of exploitation techniques or code available to calculate the temporal score. Recently discovered vulnerabilities and vulnerabilities with no available patches evidently pose a higher level of risk than vulnerabilities for which patches are available. Moreover, the availability of software updates does not ensure that all software with vulnerabilities is updated. Sometimes, users may not understand the reason behind updates and neglect them. Other times, they may slowly patch vulnerabilities as they face security-related issues. However, early updates to applications always make the system less vulnerable. With temporal Exploitability (E) and Remediation Level (RL) metrics, we calculated the temporal score, $TS_i$, using Eq. (4). Here, $TS_i$ represents the temporal score for the $i_{th}$ vulnerability, based upon the exploitability and remediation levels of the given vulnerability. The function used is:

$$TS_i = Exploitability(E_i) * RemediationLevel(R_i) \quad (4)$$

*4.4.3. Estimating the time*

More frequent vulnerabilities and delayed patch applications increase the system's risk. Following this knowledge, the proposed equation introduced a new dimension to the available standard quantitative risk model. The patch delay function is a mathematical way to capture how the risk associated with an unpatched vulnerability increases the longer a user delays applying an update. This function allows us to quantify the cumulative risk posed by deferring updates, even if users are initially unaware of the vulnerability. Here, we have utilized the exponential decay function to approximate the interval between the release of a software update and its installation in the system. This approach provides a reliable and accurate estimation of the time duration. $f_{\text{patch\_delay}}(t_{\text{patch}})$ has been used to evaluate the time the user takes to apply the patch to their system after the patch is available. The function in Eq. (5) represents the exponential progress factor of update application duration, which reflects the increase in severity level of vulnerability with delay.

$$f_{\text{patch\_delay}}(t_{\text{patch}}) = 1 - e^{-\lambda t_{\text{patch}}} \quad (5)$$

The delay is defined as the time gap between the update availability and its application to the system. Here, $\lambda$ is the rate parameter in the equation that is specified by $(1/\mu)$. Here, meu ($\mu$) is the average day for an update application, and $t_{patch}$ is the actual time that a user takes to apply the update. The exponent $1 - e^{-\lambda t_{patch}}$ determines the rate at which the function decays. As time increases, the exponent becomes more negative, causing the expression $1 - e^{-\lambda t}$ to approach zero, representing the diminishing value of the quantity being modeled.

*4.4.4. Proposed equation*

As has been discussed, delaying software updates makes the system riskier. Thus, by leveraging the CVSS score, we propose a quantitative risk evaluation equation. The base score reflects the cumulative vulnerability risk score, while the temporal score represents the risk level that changes over time. After applying Eqs. (3)–(5), the proposed formula to estimate the change in risk level is shown in Eqs. (6a) and (6b). The metrics and metric values of CVSS that we used for the equations are described in Table 1, while Table 2 represents the qualitative severity rating scale.

$$\text{Risk Score} = \sum_{i=1}^{n} \left[ BS_i + TS_i \cdot f_{\text{patch\_delay}}(t_{\text{patch}}) \right] \quad (6a)$$

From Eq. (6a), we derive Eq. (6b):

$$\text{Risk Score} = \sum_{i=1}^{n} \left[ BS_i + \left[ [E_i \cdot RL_i] \cdot f_{\text{patch\_delay}}(t_{\text{patch}}) \right] \right] \quad (6b)$$

Eq. (6b) represents the total number of vulnerabilities for which patches are available. For each vulnerability $i$, $BS_i$ is the base score, $TS_i$ is the temporal score where $E_i$ represents the exploitation value, $RL_i$ is the remediation level of the vulnerability, and function $t_{patch}$ calculates the delay in patch application. The risk score for the vulnerability will be in the form of a positive decimal number.

*4.5. Survey study on risk communication*

To assess whether our developed risk score impacts users' software update behavior, we conducted a survey study. This study aimed to determine if highlighting the risks associated with delaying updates would encourage participants to update software more proactively. Prior to the implementation of the final survey, we conducted a pilot study. The initial pilot study helped us enhance the design of our survey instrument, ascertain the clarity of the questions, and evaluate the usability and relevance of the questionnaire. Responses and feedback from pilot study participants were utilized to clarify the questions and revise the survey layout. Performing a pilot study ensured that the final survey provided reliable data on users' attitudes and behaviors toward software updates.

*4.5.1. Participant recruitment*

An online survey was distributed to the participants using Qualtics [99] to collect data. Qualtrics has been used in much previous security-related research to know more about users' perspectives [100–103]. This wide use of the software made it a potential platform for our study. The survey was designed to adopt a purposive, non-probability sampling strategy for recruiting the participants. To participate in this study, the participants were required to be at least 18 years old and have prior experience using the Windows operating system. Based on these criteria, 63 participants were initially recruited. However, 15 responses were excluded due to incompleteness, resulting in a final sample size of 48 participants. Participation in the survey was entirely voluntary, and no monetary or material compensation was provided. The study was approved by the author's institution's Institutional Review Board (IRB). We did not collect or store any form of personal information that uniquely identifies the participants when combined with other information. The researchers did not request any information that could potentially be used to breach participants' privacy or security. Therefore, the survey responses were anonymous and non-traceable.

*4.5.2. Survey design*

The survey included quantitative and qualitative questions regarding users' general behavior and opinions about software updates. At first, participants were asked some background questions for demographics analysis. These questions were related to participants' age, gender, education level, and their regular duration of computer use. Next, questions regarding specific software from different categories were posed. Questions were asked about the participants' decisions and thoughts about software updates and their knowledge of risk scores. Additionally, the survey presented participants with information on increased risk scores due to delays in updating the system.

Some questions in the survey respondents were about their experiences with specific software update notifications. We gathered information about users' opinions on security and privacy since computer security and better performance can be achieved by keeping software up-to-date. As we were focused on observing the behavioral changes of users, we utilized a widely used behavioral model, Affect-Reason-Involvement (ARI) model [104], to design the questionnaire. This ARI model helped collect users' emotional and rational appeals (e.g., annoying, negative exp) regarding the software updates. Responses for these questions were given a 5-point Likert scale rating from Never = 1 to Always = 5.

To better understand users' attitudes toward software updates, additional quantitative questions focused on different categories of software. The questionnaire contained several questions to assess the timeliness of applying software updates. As discussed previously, users often tend to delay the software update process; hence, they were asked how long they waited before applying updates for different software. The participants could choose from six options: on the same day, within a week, within two weeks, within a month, within two months or more, and never. After recording their responses. Next, the participants were simultaneously presented with vulnerability and risk-score information and asked how long they would wait before applying the update with or without the information based on three scenarios. The comparison among these three scenarios (i.e., scenario 1: without the number of vulnerability and risk score for the particular software, scenario 2: with vulnerability number for the particular software, scenario 3: with vulnerability number and risk score for not updating the software) helped us understand the users' behavioral change and concerns in applying the updates in a timely manner for different scenarios.

## 5. Results

In this Results section, we present a descriptive analysis of participants' responses and report findings for each research question (RQ1, RQ2, and RQ3). For RQ2 and RQ3, we include the null hypotheses and alternative hypotheses and discuss conclusions drawn from statistical findings.

Table 4
Study participant demographics.

| Participant type | |
|---|---|
| *Gender* | Percentage |
| *Male* | 54.17% |
| *Female* | 45.83% |
| *Other* | 0% |
| Age range | |
| Age (years) | Percentage |
| 18–25 | 35.42% |
| 26–34 | 47.92% |
| 35–55 | 14.58% |
| Preferred not to say | 2.08% |
| Education | |
| Education level | Percentage |
| Some High School | 0.00% |
| High School Diploma | 0.00% |
| Some College | 18.75% |
| Associate Degree | 0.00% |
| Bachelor's Degree | 16.67% |
| Masters Degree | 45.83% |
| PhD Degree | 16.67% |
| Other | 2.08% |
| Daily computer usage | |
| Computer use (h) | Percentage |
| 0–1 h | 0.00% |
| 1–3 h | 6.25% |
| 3–6 h | 41.67% |
| 6–8 h | 27.08% |
| 8 h or more | 25.00% |

*5.1. Participants descriptive analytics*

We recruited 63 participants for this study. Based upon our findings, the survey participants were predominantly between 26–34 years old, with more males than females (54.17% vs. 45.83% respectively). Nearly 63% of the survey participants had higher levels of education with either a master's or a Ph.D. degree. The remaining participants completed some level of education. None of the participants reported spending less than an hour a day on the computer. Table 4 represents the demographic breakdown information gained from the participants.

*5.2. Answer to RQ1: How do users' cognitive states affect their adoption of software update?*

*5.2.1. Awareness about software update*

The survey asked participants to rate their awareness of computer security and their concerns regarding software updates. The responses to software update awareness-related questions are presented in Fig. 3. The data shows that approximately 55% of showed great concerns (rated as often to always concerned) about securing their system, while around 30% of participants are only sometimes concerned. Half of the participants sometimes rely on software updates to secure their system, which is troublesome because software updates are crucial in keeping a system secure. However, users have a lack of understanding or minimal trust in software updates. Only a small percentage (10.42%) of respondents agreed on the importance of always keeping their software up to date, while merely 6.25% believed software updates might be the possible option that could save the system from being vulnerable. On the other hand, around 10% of the participants expressed skepticism toward the potential benefits of software updates in enhancing security measures. The majority of participants were less confident in their awareness and belief concerning software updates. To address this issue, software updates need to provide more information about the benefits of applying an update.

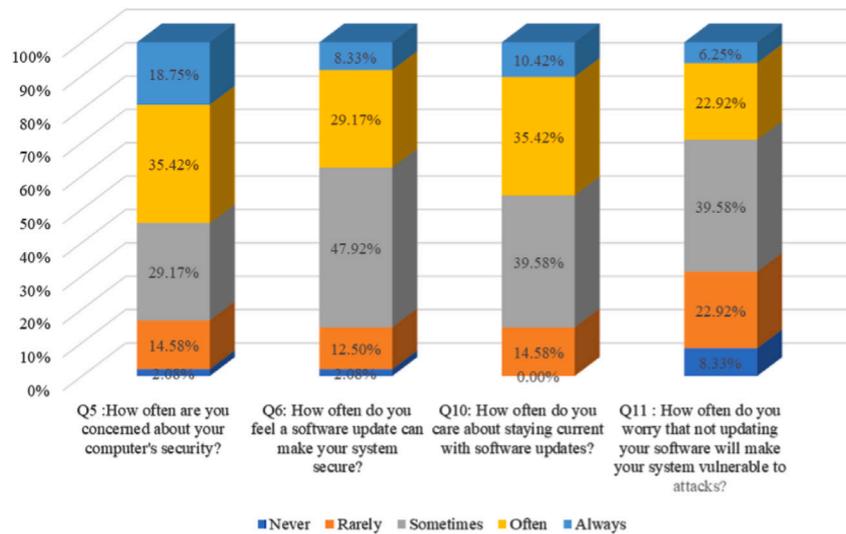

**Fig. 3.** Responses on user's awareness about software update.

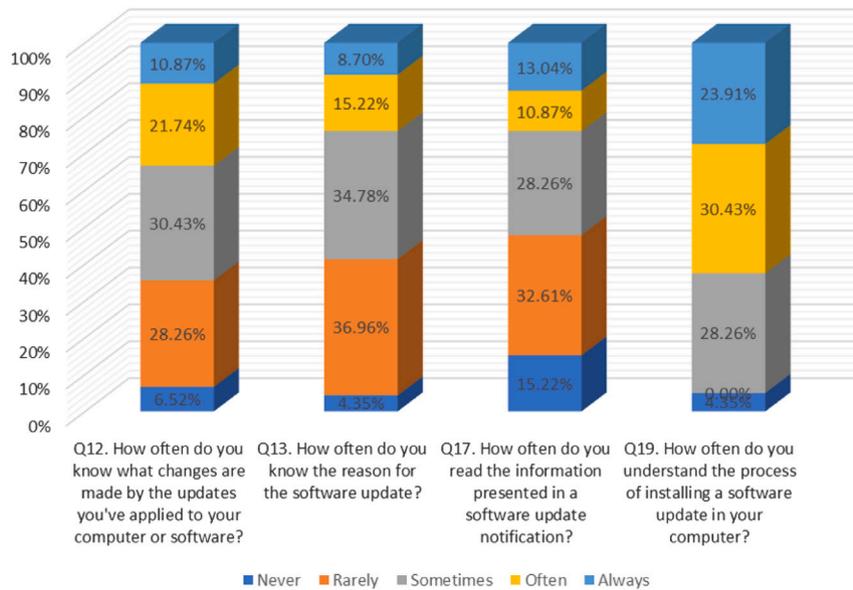

**Fig. 4.** Responses on user's knowledge about software update.

*5.2.2. Knowledge about software updates*

The prevalence of trust in software updates depends on the user's prior knowledge about the software update. Fig. 4 provides a synopsis of the participants' prior knowledge of software updates. Findings from the responses indicate that only 30.43% of users are aware of the changes that updates will make, whereas a mere 8.70% know the reasons behind the updates. The survey further revealed that almost half of the participants, approximately 50%, had a poor understanding of the installation process. Additionally, almost 60% of users scarcely read the information provided in the software update. Consequently, can be deduced that the present awareness creation tactics used to convey knowledge (or risk) about software updates are ineffective in capturing users' attention toward them.

*5.2.3. Experience with software update*

As presented in Fig. 5, From the survey it was observed that approximately 70% of the participants rarely or seldom had negative experiences in relation to software updates. Nevertheless, most users appeared hesitant to apply updates and perceived them as interruptions to their work. Notably, nearly 25% of the participants acknowledged that they often experienced hesitancy in applying any update, while approximately 37.50% of the users stated that they were sometimes hesitant, and 15% of the users reported always feeling hesitant in applying software updates. As a consequence, users tend to delay the process of updating. Vaniea et al. [55] conducted a study that revealed the tendency of users to find updates not worth the time, thus resulting in a reluctance to update. Providing the installation file along with detailed information about the new update can help to enhance user awareness. However, the survey results indicated that only about 30% of the participants considered the information in software update messages to impact their decisions, while 10% of the users reported that such information did not affect their decisions at all. Although software updates provide information about new features and previous vulnerability fixes, it has not been proven to have a significant impact on users' decisions or attitudes toward updates.

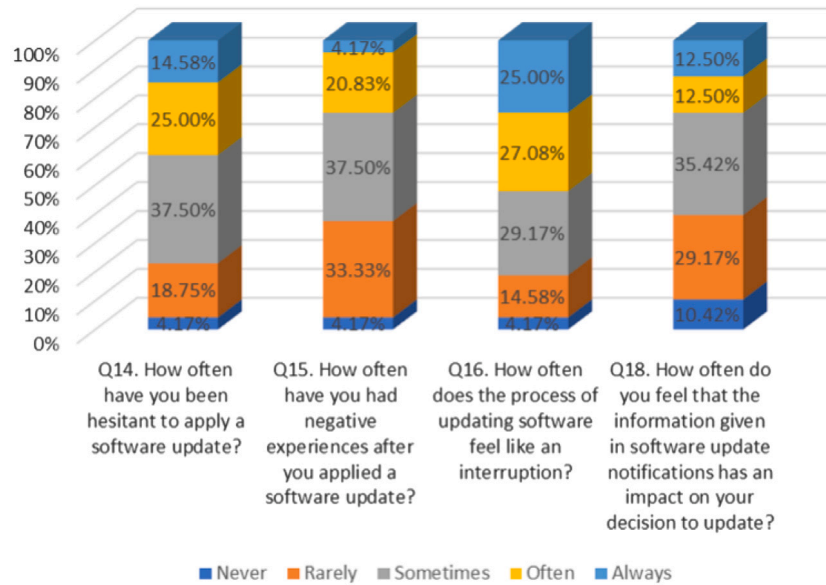

**Fig. 5.** Responses on user's experience with software update.

*5.3. Answer to RQ2: To what extent does the vulnerability and risk score information improve users' software update compliance behavior?*

Null Hypothesis ($H_0$):*Providing vulnerability and risk score information does not improve users' compliance with software updates.*

Alternative Hypothesis ($H_1$):*Providing vulnerability and risk core information improves users' compliance with software updates.*

*5.3.1. Effect of providing vulnerability number and risk score to improve user willingness to update software*

As previously mentioned, this study aims to evaluate whether providing software users with vulnerability and risk-score information affects their behavior in applying software updates. The survey questionnaire included questions asking participants how likely they were to update the listed software (list of selected software exhibited in Table 3). However, participants were presented with three different scenarios. Firstly, they were not given any vulnerability or risk information about the software. Then, in subsequent questions, participants were provided with vulnerability information, and finally, they were given the risk score information. The average score of each participant's likelihood of updating the selected 13 software was calculated using Eq. (7) in the three different scenarios mentioned: (i) before providing any information, (ii) after providing vulnerability information, and (iii) after providing vulnerability risk-score information.

$$\text{Average Score (AS)} = \frac{\sum_1^n L_s}{N} \qquad (7)$$

where AS is the average core for each participant's likelihood of applying the software update, L is the participant response for each software, and $N$ is the total number of software.

We calculated the average score for each participant's response from the three scenarios mentioned earlier. To determine if the likelihood of applying software updates significantly varies across the three scenarios, we employed the repeated measure analysis of variance (ANOVA) test. Since the same participants were evaluated under all three conditions, a within-subjects ANOVA was the appropriate choice for this design.

Before conducting the ANOVA, we assessed the data for parametric or non-parametric assumptions. To decide between parametric and non-parametric tests, we checked for the normality and homogeneity of the dataset. We used the Skewness–Kurtosis and Shapiro–Wilk tests to test data normality and Leven's test to assess data homogeneity.

The results confirmed that the data met the assumptions for parametric testing: the distributions were approximately normal, and the variances were homogeneous across conditions. Following this validation, we proceeded with the repeated measures ANOVA at a 95% confidence level. As presented in Table 5, the ANOVA revealed a statistically significant difference in participants' willingness to apply software updates across the three scenarios (F(2, 94) = 81.291, $p < 0.001$). To further assess the magnitude of this difference, we calculated the effect size using partial ($\eta^2$). The resulting value, ($\eta^2$) = 0.634, indicates that approximately 63.4% of the variance in participants' willingness to update software was explained by the type of information provided. According to Cohen's guidelines [105], this reflects a large effect size, suggesting that the informational scenarios had a substantial influence on participant behavior. Although the ANOVA test confirmed a significant effect, it did not reveal which specific scenario led to the highest willingness to update. Therefore, a post-hoc pairwise comparison was conducted to examine differences between the three scenarios. As the data was normally distributed, we used parametric post-hoc tests. The results, shown in Table 6, indicated statistically significant differences (all $p < 0.05$) for every pairwise comparison. Specifically, when comparing Scenario 1 to Scenario 2, the mean difference was −1.222, indicating greater willingness in Scenario 2. Comparing Scenario 3 to Scenario 2 yielded a mean difference of 0.384, showing that participants were even more inclined to update when risk scores were included. The largest effect was observed between Scenario 3 and Scenario 1, with a mean difference of 1.606, confirming that participants were most willing to apply the update in Scenario 3.

To summarize the findings, the participants' decision-making regarding software updates was influenced by both the vulnerability number and risk score. However, upon further analysis using pairwise comparison, it was observed that the participants were more willing and performed better in decision-making when presented with the risk score in addition to the vulnerability information rather than vulnerability information alone.

*5.3.2. Effect of vulnerability and risk score information on accelerating the software update process*

Apart from the unwillingness to apply software updates, studies also suggest that software users tend to delay the software update for various reasons (as discussed in the literature review section), making the system more vulnerable. The survey questionnaire had questions to assess whether providing software vulnerability and risk

**Table 5**

ANOVA test results.

|  | Mean | Std. deviation | N | df | F | p | Results |
|---|---|---|---|---|---|---|---|
| Scenario 1 (without information) | 2.4698 | 0.72899 | 48 |  |  |  | There are statistically |
| Scenario 2 (with vulnerability information) | 3.6919 | 0.7301 | 48 | 2 | 81.291 | <0.05 | significant differences |
| Scenario 3 (with risk-score information) | 4.076 | 0.65898 | 48 |  |  |  | among the three scenarios |

**Table 6**

Post-hoc pairwise comparison.

| Pair of scenarios | Mean difference | Std. error | p | Difference(95% CI) | | Results |
|---|---|---|---|---|---|---|
|  |  |  |  | Lower bound | Upper bound |  |
| 2 | −1.222* | 0.128 | <0.05 | −1.539 | −0.905 | Scenario 2 and 3 |
| 3 | −1.606* | 0.152 | <0.05 | −1.85 | −1.228 | demonstrated better results than |
| 1 | −0.222* | 0.128 | <0.05 | 0.905 | 1.539 | Significantly Scenario 1 |
| 3 | −0.384* | 0.111 | <0.05 | −0.660 | −0.108 | with scenario 3 exhibiting even |
| 1 | 1.606* | 0.152 | <0.05 | 1.228 | 1.985 | Greater improvement |
| 2 | −0.384* | 0.111 | <0.05 | 0.108 | 0.660 | compared to scenario 2 |

* Statistically significant

**Table 7**

ANOVA test results.

|  | Mean | Std. deviation | N | df | F | p | Results |
|---|---|---|---|---|---|---|---|
| Scenario 1 (without information) | 3.6765 | 0.96811 | 48 |  |  |  | There are statistically significant differences |
| Scenario 2 (with vulnerability information) | 2.5775 | 0.69996 | 48 | 2 | 77.38 | <0.05 | significant differences |
| Scenario 3 (with risk-score information) | 2.0413 | 0.78095 | 48 |  |  |  | among the three scenarios |

score information could improve user behavior. The participants were asked how long they waited before applying the software update for different software. The participants were given three scenarios similar to the previous analysis (Section 5.2.1). First, they were not given any vulnerability or risk score information and were asked how long they waited before applying the update. Then, participants were given the vulnerability and risk score information simultaneously and were asked the same question again. The given options were on the same day, within a week, within two weeks, within a month, within two months or more, and never. In order to conduct quantitative analysis, the possible options were converted to numerical values ranging from 1 to 6, where 1 was equivalent to on the same day, and 6 was equivalent to never. Once the participants' responses were recorded, the average score for each of the participants was calculated using Eq. (7) in a similar manner.

We calculated the average score for the amount of time participants would wait before applying a software update. We compared the average score between three different scenarios using repeated measures ANOVA analysis. However, before proceeding with the analysis, we checked the normality and homogeneity of the dataset using Skewness–Kurtosis, Shapiro–Wilk, and Leven's tests. The results showed that the dataset was normally distributed and had homogeneous variance. After confirming the normality of the dataset, we conducted a repeated measures ANOVA test to evaluate the difference in users' behavior on the provision of vulnerability and risk-score information. The results of the ANOVA test are presented in Table 7.

From Table 7, there is a significant difference in the users' delay in updating their software among the three scenarios. The mean value for scenario 1 is 3.67, indicating that users tend to wait for around 30 days before applying software updates. However, when they are provided with vulnerability information, the wait time reduces to approximately two weeks. Finally, when they are provided with risk-score information, they indicate that they would update their software within two weeks of first receiving the software update notification. Although the ANOVA test showed a significant difference among the three scenarios in the users' behavior toward software updates, it is important to conduct a pairwise comparison to determine which scenario performs the best. To achieve this objective, a post-hoc test was conducted to compare the scenarios pairwise. The results of the post-hoc analysis are shown in Table 8.

In Table 8, we can observe that all comparisons are significantly different from each other. However, the most important finding from this analysis is that participants performed better in scenario compared to scenarios 2 and 1. The difference in performance is statistically significant. In other words, when participants received risk information, they were more likely to apply the software update sooner than when they received vulnerability information or no information at all.

Furthermore, we tested if the variation in vulnerability number for each software has an impact on users' decisions. We divided the list of software into two groups: one group with a higher vulnerability number and the other group with a lower vulnerability number. Similarly, we categorized the participants' responses into two sets to determine if a higher number of vulnerabilities influences users to apply software updates more frequently. To analyze this, we conducted a paired t-test on the two groups of responses. Table 9 presents the results of the paired t-test conducted on the dataset.

In Table 9, the mean score for high-vulnerability software is 4.163, while the mean score for low-vulnerability software is 3.401. The mean difference between them is 0.761. Although this difference may appear small, the paired t-test indicates that it is statistically significant. This means that the participants are more likely to apply software updates when the software has a higher number of vulnerabilities. Conversely, they are less likely to apply the update when the software has a lower number of vulnerabilities.

To understand the impact of providing risk information to users, we utilized the proposed equation to calculate the risk score of the software due to a 30-day delay in updating. The participants were asked:

*"The risk score of Adobe Acrobat reader is 6.8 now, but if you delay the update for 30 days, then the risk score might increase to 8.05. Knowing that the delay can increase the risk level, how likely are you to apply the software update sooner?"*

As can be seen in Fig. 6, after knowing the increased risk score due to the delay in applying the software update, 56% of the participants expressed they would extremely likely apply the software update sooner than 30 days, and 29% of the participants agreed on somewhat likely.

**Table 8**
Post-hoc analysis for the delay in software update.

| Pair of scenarios | Mean difference | Std. error | p | Difference(95% CI) | | Results |
|---|---|---|---|---|---|---|
| | | | | Lower bound | Upper bound | |
| 2 | 1.099* | 0.148 | <0.05 | 0.732 | 1.466 | Scenario 2 and 3 |
| 3 | 1.635* | 0.157 | <0.05 | 1.245 | 2.025 | demonstrated significantly better results than |
| 1 | −1.099* | 0.148 | <0.05 | −1.466 | −0.732 | Scenario 1 and |
| 3 | .536* | 0.085 | <0.05 | 0.324 | 0.748 | with Scenario 3 exhibiting even |
| 1 | −1.635* | 0.157 | <0.05 | −2.025 | −1.245 | greater improvement |
| 2 | −.536* | 0.085 | <0.05 | −0.748 | −0.324 | compared to Scenario 2 |

\* Statistically significant

**Table 9**
Post-hoc analysis for the delay in software update.

| | Mean | N | Std. error mean | Mean difference | t | df | Sig | Result |
|---|---|---|---|---|---|---|---|---|
| High vulnerability software | 4.163 | 48 | 0.119 | 0.761 | 8.091 | 47 | <0.05 | Statistically significant differences |
| Low vulnerability software | 3.401 | 48 | 0.112 | | | | | |

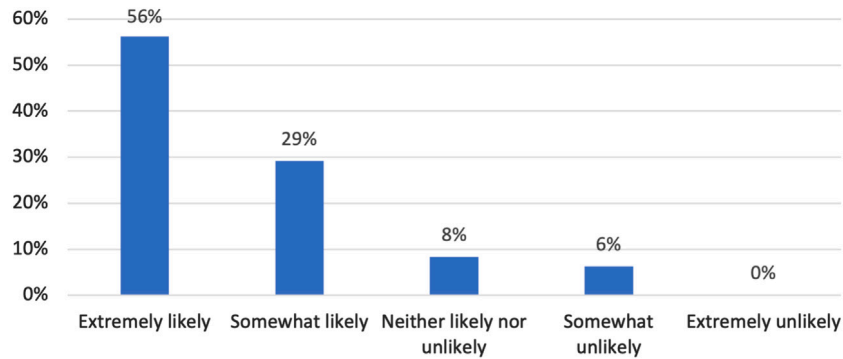

**Fig. 6.** Users' responses for applying software update after providing risk-score fo delay.

The previous analysis revealed that software users tend to defer applying software updates for around 30 days. There could be several reasons for this behavior, including busy schedules, concerns about the update's compatibility with existing software, or a lack of awareness about the consequences of delayed updates. However, our study shows that providing risk score information to software users can help change their behavior and expedite their application of software updates. Specifically, we found that nearly 80% of the participants agreed to apply software updates sooner than 30 days to mitigate the risk of being exploited due to software vulnerabilities. This finding suggests that risk score information can serve as an effective metric to persuade users to prioritize software updates and enhance security posture.

The risk score information provided to the participants was based on a thorough assessment of the software's vulnerability and the potential impact of a cyberattack. The score was communicated in a clear and concise manner, making it easy for the participants to understand the risks and benefits of promptly applying software updates. Our study underscores the significance of imparting risk score information to software users to bolster their cybersecurity preparedness and limit their vulnerability to cyber threats.

Our analyses indicate that providing vulnerability and risk score information significantly improves users' compliance with software updates. Based on our statistical analysis, there was sufficient evidence to reject the null hypothesis that providing vulnerability and risk score information does not improve users' compliance with software updates. This finding suggests that when users are informed about specific risks associated with delayed updates, they are more likely to promptly apply updates, highlighting the importance of transparent risk communication in encouraging better security practices.

*5.4. Answer to RQ3: What difference does gender make in software update decision-making?*

Null Hypothesis ($H_0$): *There is no significant difference between male and female participants in software update decision-making.*

Alternative Hypothesis ($H1$): *There is a significant difference between male and female participants in software update decision-making.*

In order to achieve the last objective of this study, we divided the dataset for each scenario into two groups based on the gender of the participants. We performed an independent t-test across the three scenarios presented in Table 10 to determine if there is any statistically significant difference between the behavior of male participants and female participants. Our findings highlighted whether gender influences the behavior of participants. As presented in Table 10, the dataset has 25 male and 23 female participants. Our previous analysis established that participants' willingness to apply software updates increases when vulnerability and risk-score information is provided. However, based on the gender-based differences, post hoc analysis, with a *p*-value greater than 0.05, revealed no significant difference in software update decision-making between male and female participants. For each of the three scenarios, the mean score for both gender groups is closely aligned, and the p-val e exceeds 0.05 in all cases. This result suggests that gender does not play a statistically significant role in influencing the decision to update software in this sample. Based on the results of the statistical test, there was insufficient evidence to reject the null hypothesis that there is no significant difference between males and female participants in software update decision-making.

**Table 10**
Post-hoc analysis for the delay in software update.

| Scenario | Gender | N | Mean | Std. deviation | F | *p*-value | Result |
|---|---|---|---|---|---|---|---|
| Scenario 1 | Male | 25 | 2.5568 | 0.69029 | 0.542 | 0.466 | |
| | Female | 23 | 2.3752 | 0.77301 | | | Statistically |
| Scenario 2 | Male | 25 | 3.6948 | 0.667 | 0.95 | 0.335 | No significant |
| | Female | 23 | 3.6887 | 0.80834 | | | differences |
| Scenario 3 | Male | 25 | 4.0404 | 0.61837 | 0.426 | 0.517 | |
| | Female | 23 | 4.1148 | 0.71243 | | | |

## 6. Discussion

### 6.1. Users' willingness to update

Our study demonstrates that users' willingness to apply software updates increases significantly when they are provided with clear, structured risk information. Scenario containing vulnerability counts showed greater willingness to update compared to a baseline scenario with minimal information. More notably, willingness further increased when participants were shown both the vulnerability count and a dynamic risk score that communicated how risk escalates with delay. This suggests that users are not inherently averse to updates but rather respond to the salience and framing of security information. These findings align with previous work emphasizing the role of cognitive framing and risk perception in security decision-making. In addition, the study revealed that users found visual cues, such as rising risk scores over time especially persuasive. This highlights the need for software vendors to design update mechanisms that can enhance user responsiveness and increase compliance toward software updates.

### 6.2. Impact of gender in security risk assessment

The finding suggests that there was no statistically significant difference between male and female users regarding security concerns and willingness to update software was partially unexpected, given the mixed results reported in prior literature. Several earlier studies have observed gender-based differences in cybersecurity behavior, for instance, women were found to be more concerned about security and privacy in some contexts [85–87], yet other studies reported women as more susceptible to phishing or less likely to adopt technical security measures [78,89,90]. Given this, we considered the possibility of gender playing a role in update decision-making. However, our study found no significant differences across all three experimental scenarios (as shown in Table 10). One possible explanation is that when users, regardless of gender, if users are provided with clear and quantifiable risk information (e.g., vulnerability numbers and risk scores), their decisions converge. This suggests that structured risk communication may help mitigate behavioral disparities typically attributed to demographic factors. Moreover, our sample consisted of relatively well-educated participants, with 63% holding master's or PhD degrees, which might have contributed to more homogeneous security attitudes and behaviors across genders. We have acknowledged this in our limitations section and plan to explore broader demographic diversity in future studies to better understand this dynamic.

### 6.3. Design recommendations for software update notifications

Based on our findings, there was a significant improvement in user compliance when risk-related information was provided. We propose the following design recommendations for software developers and manufacturers to improve the effectiveness of software update notifications.

- Display Quantified Risk Information: Clearly communicating CVSS-based vulnerability scores provides users with tangible, quantifiable information to assess risk accurately. Explicitly including numeric risk ratings (e.g., "Risk score: 8.0 – High Severity") could help users better comprehend the urgency of security-related actions.
- Use Visual Risk Cues: Leverage visual elements such as color-coded severity scales (e.g., red for critical, yellow for moderate) to enhance risk salience. Studies in usable security have shown that visual risk indicators improve user comprehension and compliance with warnings [46,106,107].
- Highlight Consequences of Delay: Briefly articulate what might happen if the update is not applied, for example, integrating warning *This vulnerability can allow remote code execution* in update notifications. Explaining potential consequences could enhance user motivation to comply with security advice.
- Provide Update Time Estimates: Displaying the estimated time required for the update reduces perceived disruption and increases the likelihood of compliance [55]. Users are more likely to update when they feel they can plan for the interruption.
- Avoid Ambiguity in Language: Use specific, transparent language rather than generic statements like *important update available*. Users often mistrust vague notifications or dismiss them if the message lacks clarity about purpose or risk and lack of information [108]
- Leverage Risk Progression Messaging: Present users with information showing how risk escalates over time with delay. For example, "Risk Score now: 6.8; in 30 days: 8.0".

By embedding these elements into update notifications, developers may increase users' motivation to apply updates promptly, thereby improving system-wide security compliance.

## 7. Conclusion

Maintaining software security requires timely program updates, yet user noncompliance remains a persistent challenge. In this study, we examined the psychological and informational factors influencing users' decisions to apply updates. Through a structured survey, we analyzed user attributes specifically, awareness, experience, and knowledge and their relationship to update behavior. Our findings reveal that perceptual and informational gaps are frequently the cause of delays in applying software updates. To address this, we proposed a risk assessment model that quantifies the severity of software vulnerabilities by leveraging attributes from the National Vulnerability Database (NVD). Specifically, our model incorporates a patch delay function to capture how the risk increases over time if updates are postponed. We validated the effectiveness of this model through a user study that presented participants with three scenarios: baseline, vulnerability count, and vulnerability + risk score (with delay). The results showed that providing quantified vulnerability and risk information significantly increased users' willingness to apply software updates, with the strongest compliance observed when users were shown how risk escalates over time. This indicates that risk transparency and temporal framing are effective mechanisms for motivating timely update behavior. Furthermore, our analysis found no statistically significant gender-based differences in update decisions, suggesting that structured and quantifiable risk information may mitigate demographic variance in security-related choices. In summary, this work contributes a risk-based behavioral framework for understanding and improving software update compliance. It provides both a computational model and empirical evidence supporting the design of more effective, information-rich update notifications. Future work will explore real-time deployment challenges and extend the model to diverse platforms and user populations.

## 8. Limitation and future work

This study addresses the Windows platform exclusively while also possessing the potential to be extrapolated to other platforms, such as Mac and Android, in subsequent experimental analyses. The researchers collected data from various software domains to determine participants' average score, which represents their inclination to apply software updates. In future inquiries, researchers could scrutinize user behavior with respect to distinct software domains, such as Windows updates, antivirus software, multimedia software, and others. Our findings are based on a moderate sample size (n = 48), primarily composed of well-educated participants with relatively high computer proficiency. This demographic skew may restrict generalizability. Additionally, the demographic analysis considered only gender, and no significant gender-based differences were observed. However, broader demographic factors such as age, educational background, and technical expertise may influence software update behavior and should be incorporated in future studies. Our demographic analysis focused only on gender (male and female). While no significant gender-based differences were found, future research could investigate the role of other psychosocial and behavioral factors, including privacy attitudes, cognitive load, and risk perception. Finally, our study measured self-reported behavioral intention, which may not fully reflect actual user behavior in real-world settings. To strengthen real time validity, we plan to conduct controlled field experiments or longitudinal studies that observe real update behavior in response to different notification strategies and risk presentations. In addition, we also aim to explore the challenges in real-time application and attack scenarios.

## CRediT authorship contribution statement

**Mahzabin Tamanna:** Writing – review & editing, Writing – original draft, Visualization, Validation, Software, Resources, Methodology, Investigation, Formal analysis, Data curation, Conceptualization. **Mohd Anwar:** Writing – review & editing, Supervision, Resources, Project administration, Funding acquisition. **Joseph D.W. Stephens:** Writing – review & editing, Project administration.

## Declaration of competing interest

The authors declare the following financial interests/personal relationships which may be considered as potential competing interests: Mahzabin Tamanna reports financial support was provided by National Science Foundation. If there are other authors, they declare that they have no known competing financial interests or personal relationships that could have appeared to influence the work reported in this paper.

## Acknowledgments

We express our sincere gratitude to the National Science Foundation (NSF), grant 2007662, for providing funding for this research. Their support has been vital in advancing our study and contributing to the scientific community. We would also like to thank the North Carolina A&T State University Human-Centered AI research group for their valuable support in this paper.

## Data availability

Data will be made available on request.

## References

[1] Salini P, Kanmani S. Effectiveness and performance analysis of model-oriented security requirements engineering to elicit security requirements: a systematic solution for developing secure software systems. Int J Inf Secur 2016;15:319–34.

[2] Magnanini F, Ferretti L, Colajanni M. Scalable, confidential and survivable software updates. IEEE Trans Parallel Distrib Syst 2021;33(1):176–91.

[3] Li J, Reiher PL, Popek GJ. Resilient self-organizing overlay networks for security update delivery. IEEE J Sel Areas Commun 2004;22(1):189–202.

[4] Johansen H, Johansen D, van Renesse R. Firepatch: Secure and time-critical dissemination of software patches. In: IFIP international information security conference. Springer; 2007, p. 373–84.

[5] Ambrosin M, Busold C, Conti M, Sadeghi A-R, Schunter M. Updaticator: Updating billions of devices by an efficient, scalable and secure software update distribution over untrusted cache-enabled networks. In: Computer security-ESORICS 2014: 19th European symposium on research in computer security, wroclaw, Poland, September 7-11, 2014. proceedings, part i 19. Springer; 2014, p. 76–93.

[6] Yohan A, Lo N-W. FOTB: a secure blockchain-based firmware update framework for IoT environment. Int J Inf Secur 2020;19(3):257–78.

[7] Why one in four downloads still has a Log4j vulnerability https://accelerationeconomy.com/cybersecurity/why-one-in-four-downloads-still-has-a-log4j-vulnerability/#:~:text=One%20year%20ago%2C%20the%20Log4j,that%20use%20Java%20were%20affected.

[8] National vulnerability database https://nvd.nist.gov/.

[9] What is the Log4j vulnerability?,.

[10] Software patching statistics: Common practices and vulnerabilitieshttps://heimdalsecurity.com/blog/software-patching-statistics-practicesvulnerabilities/.

[11] Top 30 targeted high risk vulnerabilities | CISA," Cybersecurity and Infrastructure Security Agency https://www.cisa.gov/news-events/alerts/2015/04/29/top-30-targeted-high-risk-vulnerabilities.

[12] Forget A, Pearman S, Thomas J, Acquisti A, Christin N, Cranor LF, Egelman S, Harbach M, Telang R. Do or do not, there is no try: user engagement may not improve security outcomes. In: Twelfth symposium on usable privacy and security (SOUPS 2016). 2016, p. 97–111.

[13] Cranor LF. A framework for reasoning about the human in the loop. 2008.

[14] Anderson CL, Agarwal R. Practicing safe computing: A multimethod empirical examination of home computer user security behavioral intentions. MIS Q 2010;613–43.

[15] Khan M, Bi Z, Copeland JA. Software updates as a security metric: Passive identification of update trends and effect on machine infection. In: MILCOM 2012-2012 IEEE military communications conference. IEEE; 2012, p. 1–6.

[16] Furnell SM, Jusoh A, Katsabas D. The challenges of understanding and using security: A survey of end-users. Comput Secur 2006;25(1):27–35.

[17] Stanton JM, Stam KR, Mastrangelo P, Jolton J. Analysis of end user security behaviors. Comput Secur 2005;24(2):124–33.

[18] What is cybersecurity and why is it important? | Accenture, Accenture Security and Ponemon Institute https://www.accenture.com/us-en/insights/cyber-security-index.

[19] Alshammari NO, Mylonas A, Sedky M, Champion J, Bauer C. Exploring the adoption of physical security controls in smartphones. In: Human aspects of information security, privacy, and trust: third international conference, HAS 2015, held as part of HCI international 2015, los angeles, CA, USA, August 2-7, 2015. proceedings 3. Springer; 2015, p. 287–98.

[20] Rajivan P, Moriano P, Kelley T, Camp LJ. Factors in an end user security expertise instrument. Inf Comput Secur 2017;25(2):190–205.

[21] Microsoft Security Intelligence Report https://download.microsoft.com).

[22] Patching Could Have Stopped Most Breaches, Study Finds. 2016, https://www.eweek.com/security/software-patches-could-prevent-most-breaches-study-finds/.

[23] Vulnerabilities on the corporate network perimeter. 2020, https://www.ptsecurity.com/ww-en/analytics/vulnerabilities-corporate-networks-2020.

[24] Beres Y, Griffin J, Shiu S, Heitman M, Markle D, Ventura P. Analysing the performance of security solutions to reduce vulnerability exposure window. In: 2008 annual computer security applications conference. ACSAC, IEEE; 2008, p. 33–42.

[25] Altinkemer K, Rees J, Sridhar S. Vulnerabilities and patches of open source software: An empirical study. J Inf Syst Secur 2008;4(2):3–25.

[26] Arora A, Krishnan R, Telang R, Yang Y. An empirical analysis of software vendors' patch release behavior: impact of vulnerability disclosure. Inf Syst Res 2010;21(1):115–32.

[27] Frei S. Security econometrics: The dynamics of (in) security. vol. 93, ETH Zurich; 2009.

[28] Joh H, Malaiya YK. Defining and assessing quantitative security risk measures using vulnerability lifecycle and cvss metrics. In: The 2011 international conference on security and management (sam). Citeseer; 2011, p. 10–6.
[29] Nosek BA, Banaji MR, Greenwald AG. Harvesting implicit group attitudes and beliefs from a demonstration web site.. Group Dyn: Theory, Res Pr 2002;6(1):101.
[30] McGill T, Thompson N. Exploring potential gender differences in information security and privacy. Inf Comput Secur 2021;29(5):850–65.
[31] Perrig A, Szewczyk R, Wen V, Culler D, Tygar J. SPINS: Security protocols for sensor networks. In: Proceedings of the 7th annual international conference on mobile computing and networking. 2001, p. 189–99.
[32] Wang HJ, Guo C, Simon DR, Zugenmaier A. Shield: Vulnerability-driven network filters for preventing known vulnerability exploits. In: Proceedings of the 2004 conference on applications, technologies, architectures, and protocols for computer communications. 2004, p. 193–204.
[33] Ajzen I. Nature and operation of attitudes. Annu Rev Psychol 2001;52(1):27–58.
[34] Ajzen I, Fishbein M. Attitudes and the attitude-behavior relation: Reasoned and automatic processes. Eur Rev Soc Psychol 2000;11(1):1–33.
[35] Common Vulnerability Scoring System SIG https://www.first.org/cvss/.
[36] National Vulnerability Database https://nvd.nist.gov/vuln/search/statistics/form_type=Basic&results_type=statistics&search_type=all&isCpeNameSearch=false.
[37] Mell P, Scarfone K, Romanosky S, et al. A complete guide to the common vulnerability scoring system version 2.0. In: Published by FIRST-forum of incident response and security teams. vol. 1, 2007, p. 23.
[38] Yıldırım M, Mackie I. Encouraging users to improve password security and memorability. Int J Inf Secur 2019;18:741–59.
[39] Christin N, Egelman S, Vidas T, Grossklags J. It's all about the benjamins: An empirical study on incentivizing users to ignore security advice. In: Financial cryptography and data security: 15th international conference, FC 2011, gros islet, st. lucia, February 28-March 4, 2011, revised selected papers 15. Springer; 2012, p. 16–30.
[40] Vitale F, McGrenere J, Tabard A, Beaudouin-Lafon M, Mackay WE. High costs and small benefits: A field study of how users experience operating system upgrades. In: Proceedings of the 2017 CHI conference on human factors in computing systems. 2017, p. 4242–53.
[41] Faklaris C, Dabbish LA, Hong JI. A {self-report} measure of {end-user} security attitudes ({{{{sa-6}}}}). In: Fifteenth symposium on usable privacy and security (SOUPS 2019). 2019, p. 61–77.
[42] Bryant P, Furnell S, Phippen A. Improving protection and security awareness amongst home users. Adv Networks, Comput Commun 2008;4:182.
[43] Möller A, Michahelles F, Diewald S, Roalter L, Kranz M. Update behavior in app markets and security implications: A case study in google play. In: Research in the large, LARGE 3.0: 21/09/2012-21/09/2012. 2012, p. 3–6.
[44] Nicholson J, Coventry L, Briggs P. Introducing the cybersurvival task: Assessing and addressing staff beliefs about effective cyber protection. In: Fourteenth symposium on usable privacy and security (SOUPS 2018). 2018, p. 443–57.
[45] Vaniea KE, Rader E, Wash R. Betrayed by updates: how negative experiences affect future security. In: Proceedings of the SIGCHI conference on human factors in computing systems. 2014, p. 2671–4.
[46] Fagan M, Khan MMH, Nguyen N. How does this message make you feel? A study of user perspectives on software update/warning message design. Human-Centric Comput Inf Sci 2015;5(1):1–26.
[47] Mathur A, Chetty M. Impact of user characteristics on attitudes towards automatic mobile application updates. In: Thirteenth symposium on usable privacy and security (SOUPS 2017). 2017, p. 175–93.
[48] Chowdhury N, Gkioulos V. A personalized learning theory-based cyber-security training exercise. Int J Inf Secur 2023;1–16.
[49] Angafor GN, Yevseyeva I, Maglaras L. Securing the remote office: reducing cyber risks to remote working through regular security awareness education campaigns. Int J Inf Secur 2024;1–15.
[50] Katsantonis M, Manikas A, Mavridis I, Gritzalis D. Cyber range design framework for cyber security education and training. Int J Inf Secur 2023;1–23.
[51] Ekstedt M, Afzal Z, Mukherjee P, Hacks S, Lagerström R. Yet another cybersecurity risk assessment framework. Int J Inf Secur 2023;22(6):1713–29.
[52] Wangen G, Hallstensen C, Snekkenes E. A framework for estimating information security risk assessment method completeness: Core unified risk framework, CURF. Int J Inf Secur 2018;17:681–99.
[53] Yamin MM, Katt B. Modeling and executing cyber security exercise scenarios in cyber ranges. Comput Secur 2022;116:102635.
[54] Shamala P, Ahmad R, Yusoff M. A conceptual framework of info structure for information security risk assessment (ISRA). J Inf Secur Appl 2013;18(1):45–52.
[55] Vaniea K, Rashidi Y. Tales of software updates: The process of updating software. In: Proceedings of the 2016 chi conference on human factors in computing systems. 2016, p. 3215–26.
[56] Fagan M, Khan MMH, Buck R. A study of users' experiences and beliefs about software update messages. Comput Hum Behav 2015;51:504–19.
[57] Egelman S, Cranor LF, Hong J. You've been warned: an empirical study of the effectiveness of web browser phishing warnings. In: Proceedings of the SIGCHI conference on human factors in computing systems. 2008, p. 1065–74.
[58] Howe AE, Ray I, Roberts M, Urbanska M, Byrne Z. The psychology of security for the home computer user. In: 2012 IEEE symposium on security and privacy. IEEE; 2012, p. 209–23.
[59] Asgharpour F, Liu D, Camp LJ. Mental models of security risks. In: Financial cryptography and data security: 11th international conference, FC 2007, and 1st international workshop on usable security, USeC 2007, scarborough, trinidad and tobago, February 12-16, 2007. revised selected papers 11. Springer; 2007, p. 367–77.
[60] Wash R. Folk models of home computer security. In: Proceedings of the sixth symposium on usable privacy and security. 2010, p. 1–16.
[61] Dourish P, Grinter RE, Delgado De La Flor J, Joseph M. Security in the wild: user strategies for managing security as an everyday, practical problem. Pers Ubiquitous Comput 2004;8:391–401.
[62] Ion I, Reeder R, Consolvo S. {"... No} one can hack my {Mind"}: Comparing expert and {Non-Expert} security practices. In: Eleventh symposium on usable privacy and security (SOUPS 2015). 2015, p. 327–46.
[63] Tiefenau C, Häring M, Krombholz K, Von Zezschwitz E. Security, availability, and multiple information sources: Exploring update behavior of system administrators. In: Sixteenth symposium on usable privacy and security (SOUPS 2020). 2020, p. 239–58.
[64] Bellissimo A, Burgess J, Fu K. Secure software updates: Disappointments and new challenges.. In: HotSec. 2006.
[65] Rosen S, Qian Z, Mao ZM. Appprofiler: a flexible method of exposing privacy-related behavior in android applications to end users. In: Proceedings of the third ACM conference on data and application security and privacy. 2013, p. 221–32.
[66] Fassl M, Neumayr M, Schedler O, Krombholz K. Transferring update behavior from smartphones to smart consumer devices. In: European symposium on research in computer security. Springer; 2021, p. 357–83.
[67] Haney JM, Furman SM. User perceptions and experiences with smart home updates. In: 2023 IEEE symposium on security and privacy. SP, IEEE; 2023, p. 2867–84.
[68] Sarabi A, Zhu Z, Xiao C, Liu M, Dumitraş T. Patch me if you can: A study on the effects of individual user behavior on the end-host vulnerability state. In: Passive and active measurement: 18th international conference, PAM 2017, sydney, NSW, Australia, March 30-31, 2017, proceedings 18. Springer; 2017, p. 113–25.
[69] Edwards WK, Poole ES, Stoll J. Security automation considered harmful? In: Proceedings of the 2007 workshop on new security paradigms. 2008, p. 33–42.
[70] Duebendorfer T, Frei S. Why silent updates boost security. ETH Zurich, Tech. Rep, Vol. 302, 2009, p. 98.
[71] Wash R, Rader E, Vaniea K, Rizor M. Out of the loop: How automated software updates cause unintended security consequences. In: 10th symposium on usable privacy and security (SOUPS 2014). 2014, p. 89–104.
[72] Tian Y, Liu B, Dai W, Ur B, Tague P, Cranor LF. Supporting privacy-conscious app update decisions with user reviews. In: Proceedings of the 5th annual ACM cCS workshop on security and privacy in smartphones and mobile devices. 2015, p. 51–61.
[73] Frik A, Egelman S, Harbach M, Malkin N, Peer E. Better late (r) than never: increasing cyber-security compliance by reducing present bias. In: Symposium on usable privacy and security. 2018, p. 12–4.
[74] Tian Y, Liu B, Dai W, Cranor LF, Ur B. Study on user's attitude and behavior towards android application update notification. CA: Usenix, Menlo Park,; 2014.
[75] Sheng S, Holbrook M, Kumaraguru P, Cranor LF, Downs J. Who falls for phish? A demographic analysis of phishing susceptibility and effectiveness of interventions. In: Proceedings of the SIGCHI conference on human factors in computing systems. 2010, p. 373–82.
[76] Connolly LY, Lang M, Gathegi J, Tygar JD. The effect of organisational culture on employee security behaviour: A qualitative study.. In: HAISA. 2016, p. 33–44.
[77] Gratian M, Bandi S, Cukier M, Dykstra J, Ginther A. Correlating human traits and cyber security behavior intentions. Comput Secur 2018;73:345–58.
[78] Anwar M, He W, Ash I, Yuan X, Li L, Xu L. Gender difference and employees' cybersecurity behaviors. Comput Hum Behav 2017;69:437–43.
[79] Isaksson C. Fighting for gender equality: Why security sector actors must combat sexual and gender-based violence. In: The fletcher forum of world affairs. JSTOR; 2014, p. 49–72.
[80] Detraz N. International security and gender. John Wiley & Sons; 2013.
[81] Broos A. Gender and information and communication technologies (ICT) anxiety: Male self-assurance and female hesitation. CyberPsychology & Behav 2005;8(1):21–31.
[82] He J, Freeman LA. Are men more technology-oriented than women? The role of gender on the development of general computer self-efficacy of college students. J Inf Syst Educ 2010;21(2):203–12.
[83] Venkatesh V, Morris MG. Why don't men ever stop to ask for directions? Gender, social influence, and their role in technology acceptance and usage behavior. MIS Q 2000;115–39.
[84] Mohamed N, Ahmad IH. Information privacy concerns, antecedents and privacy measure use in social networking sites: Evidence from Malaysia. Comput Hum Behav 2012;28(6):2366–75.

[85] Laric MV, Pitta DA, Katsanis LP. Consumer concerns for healthcare information privacy: A comparison of US and Canadian perspectives. Res Heal Financ Manag 2009;12(1):93.
[86] Hoy MG, Milne G. Gender differences in privacy-related measures for young adult facebook users. J Interact Advert 2010;10(2):28–45.
[87] Pattinson M, Butavicius M, Lillie M, Ciccarello B, Parsons K, Calic D, McCormac A. Matching training to individual learning styles improves information security awareness. Inf Comput Secur 2020;28(1):1–14.
[88] Jagatic TN, Johnson NA, Jakobsson M, Menczer F. Social phishing. Commun ACM 2007;50(10):94–100.
[89] Milne GR, Labrecque LI, Cromer C. Toward an understanding of the online consumer's risky behavior and protection practices. J Consum Aff 2009;43(3):449–73.
[90] Tripathi A, Singh UK. On prioritization of vulnerability categories based on CVSS scores. In: 2011 6th international conference on computer sciences and convergence information technology. ICCIT, IEEE; 2011, p. 692–7.
[91] Houmb SH, Franqueira VN. Estimating ToE risk level using CVSS. In: 2009 international conference on availability, reliability and security. IEEE; 2009, p. 718–25.
[92] Kekül H, Ergen B, Arslan H. Estimating vulnerability metrics with word embedding and multiclass classification methods. Int J Inf Secur 2024;23(1):247–70.
[93] Aksu MU, Dilek MH, Tatlı Eİ, Bicakci K, Dirik HI, Demirezen MU, Aykır T. A quantitative CVSS-based cyber security risk assessment methodology for IT systems. In: 2017 international carnahan conference on security technology. ICCST, IEEE; 2017, p. 1–8.
[94] Lenkala SR, Shetty S, Xiong K. Security risk assessment of cloud carrier. In: 2013 13th IEEE/ACM international symposium on cluster, cloud, and grid computing. IEEE; 2013, p. 442–9.
[95] Sawilla R, Burrell C. Metrics-based computer network defence decision support. In: Proceedings of the NATO RTO iST-091 symposium on information assurance and cyber defence. 2010, p. 1–22.
[96] Anwar A, Abusnaina A, Chen S, Li F, Mohaisen D. Cleaning the NVD: Comprehensive quality assessment, improvements, and analyses. IEEE Trans Dependable Secur Comput 2021;19(6):4255–69.
[97] Fruhwirth C, Mannisto T. Improving CVSS-based vulnerability prioritization and response with context information. In: 2009 3rd international symposium on empirical software engineering and measurement. IEEE; 2009, p. 535–44.
[98] Zhang S, Cai M, Zhang M, Zhao L, de Carnavalet Xd. The flaw within: Identifying CVSS score discrepancies in the NVD. In: 2023 IEEE international conference on cloud computing technology and science (cloudCom). IEEE; 2023, p. 185–92.
[99] https://www.qualtrics.com/ https://www.qualtrics.com/.
[100] Redmiles EM, Kross S, Mazurek ML. How i learned to be secure: a census-representative survey of security advice sources and behavior. In: Proceedings of the 2016 ACM SIGSAC conference on computer and communications security. 2016, p. 666–77.
[101] Danilova A, Naiakshina A, Smith M. One size does not fit all: a grounded theory and online survey study of developer preferences for security warning types. In: Proceedings of the ACM/IEEE 42nd international conference on software engineering. 2020, p. 136–48.
[102] Wash R, Rader E. Too much knowledge? security beliefs and protective behaviors among united states internet users. In: Eleventh symposium on usable privacy and security (SOUPS 2015). 2015, p. 309–25.
[103] Albrechtsen E. A qualitative study of users' view on information security. Comput Secur 2007;26(4):276–89.
[104] Buck R, Anderson E, Chaudhuri A, Ray I. Emotion and reason in persuasion: Applying the ARI model and the casc scale. J Bus Res 2004;57(6):647–56.
[105] Cohen J. Statistical power analysis for the behavioral sciences. routledge; 2013.
[106] Sun X, Jin J, Yang Y, Pan Y. Telling the "bad" to motivate your users to update: Evidence from behavioral and ERP studies. Comput Hum Behav 2024;153:108078.
[107] Darejeh A, Singh D. A review on user interface design principles to increase software usability for users with less computer literacy. J Comput Sci 2013;9(11):1443.
[108] Mathur A, Engel J, Sobti S, Chang V, Chetty M. " they keep coming back like zombies": Improving software updating interfaces. In: Twelfth symposium on usable privacy and security (SOUPS 2016). 2016, p. 43–58.